\documentclass[a4paper]{article}
\usepackage[dvips]{epsfig}
\addtocounter{secnumdepth}{1}
\newcommand{\dd}{\text{d}}
\newcommand{\ee}{\text{e}}
\newcommand{\ii}{\text{i}}
\newcommand{\ji}{\text{\footnotesize i}}
\newcommand{\p}{\partial}

\newcommand{\Ind}{\scriptsize\mbox}

\newcommand{\half}{\text{$\frac{1}{2}$}}
\newcommand{\idd}{\!\!\dd}
\newcommand{\sqtl}{\sqrt{2\lambda}}
\newcommand{\iint}{\!\!\int}
\newcommand{\text}{\mbox}

\def\gsim{\:\raisebox{-0.5ex}{$\stackrel{\textstyle>}{\sim}$}\:}

\begin{document}

\thispagestyle{empty}

\title{Imaginary noise 
and parity conservation\\
in the reaction $A+A\rightleftharpoons 0$}
 
\author{O. Deloubri\`ere$^1$,\, 
L. Frachebourg$^2$,\, H.J. Hilhorst$^1$,\, and K. Kitahara$^3$
\\[3mm]
{\small
$^1$Laboratoire de Physique Th\'eorique, B\^atiment 210}\\
{\small
Universit\'e de Paris-Sud, 91405 Orsay cedex, France}\\
{\small 
$^2$ 11 rue Fendt, 1201 Gen\`eve, Switzerland}\\
{\small
$^3$Division of Natural Sciences, International Christian University}\\
{\small
10-2 Osawa 3-chome, Mitaka-shi, Tokyo 181-8585, Japan}\\}

\maketitle
\vspace{10mm}
\begin{small}
\begin{abstract}
\noindent

The master equation for 
the reversible reaction $A+A\rightleftharpoons 0$ is considered
in Poisson representation, where
it is equivalent to a Langevin equation with imaginary noise
for a complex stochastic variable $\phi$. 
Such Langevin equations appear quite generally in
field-theoretic treatments of reaction--diffusion problems.
For this example we study the probability flow 
in the complex $\phi$ plane both analytically and by simulation.
We show that this flow has various curious features that 
must be expected to occur similarly in other 
Langevin equations associated with reaction--diffusion problems.
\end{abstract}
\end{small}
\vspace{70mm}

\noindent LPT ORSAY 01/74\\
{\small Laboratoire associ\'e au Centre National de la
Recherche Scientifique - UMR 8627}
\newpage
\section{Introduction} 
\label{secintroduction}

Let there be a reacting chemical system with diffusing
species $A, B,\ldots$
By a ``chemical'' master equation is meant one which
governs the time evolution of the
probability distribution 
of the particle numbers of these species. 
Under certain conditions such an equation is equivalent, via a
``Poisson representation,'' to a
Langevin equation with imaginary noise for space and time dependent
fields $\phi_A(x,t),\, \phi_B(x,t),\ldots$ 
This equivalence was first shown
by Gardiner and coworkers (see \cite{Gardiner}) and 
has reappeared \cite{Cardy} in 
field-theoretic treatments of reaction--diffusion problems 
as an outcome of the second-quantized formalism. 
Imaginary noise is typically (although not only) due 
to reaction processes whose reactants include two or more
particles of the same species; it is a mathematical tool 
that helps in an elegant way to keep
track of evolving probability distributions.

It is of interest to investigate such imaginary noise
Langevin equations more closely.
The motivation comes, for one part, from the intrinsic interest of a
relatively unexplored type of equation, and for another part,
from the hope that perhaps Monte Carlo simulation of 
a reaction--diffusion system in this new representation
could turn out to be efficient. This latter idea is reinforced by
the fact that the stochastic diffusive motion 
of the reacting particles becomes
{\it deterministic} in the Poisson representation: The Langevin
noise is exclusively due to the reaction processes.
 
One may therefore learn about
imaginary noise even in the absence of diffusion. 
In this note we present a case study in which,
for simplicity and in order to fully control all aspects of the
problem, we will disregard diffusion except for a brief mention
at the end.
We will examine the single-species
reversible reaction process
\begin{equation}
A+A\rightleftharpoons 0
\label{2Ato0}
\end{equation}
where the symbol ``0'' may represent an inert species.
Its only parameter is the ratio $\lambda$ of the forward $(0\to 2A)$
to backward $(2A\to 0)$ reaction rates. 

Several questions of interest may be investigated on this example.
In general, reactions may or may not
conserve the {\it parity} of the total particle number, with
ensuing consequences for 
the critical behavior of a system
(for an enlightening recent discussion see \cite{PHK}).
Hence one natural and nontrivial question here
is how the parity conservaton of
reaction (\ref{2Ato0}) is
reflected in the properties of the imaginary noise Langevin equation.
\vspace{2mm}

In its usual particle number representation the reaction (\ref{2Ato0}) 
is of course fully understood and  trivial. However, the 
corresponding Langevin equation, which is a stochastic ODE for a single
function (``field'') $\phi(t)$, has many curious features. These
will be the focus of our interest here.
We will see that not all of these features
necessarily have their counterparts in physical properties,
and will try to identify the role of each.
\vspace{2mm}

We mention some related work. 
The usual interest in the literature has been in
the case $\lambda=0$ (only particle annihilation) in the presence of
diffusion (see {\it e.g.} Lee \cite{Lee} and references therein); 
the main object of study then is
the exponent of the power law decay to the 
zero density state. 
The two coupled Langevin equations 
for the reaction $A+A\rightleftharpoons C$ 
and $A+B\rightleftharpoons C$ with diffusion
were studied in detail by Rey and Cardy \cite{ReyCardy}.
They are interested in  
how the $C$ particle densities approaches its equilibrium values
for large times. 
Howard and T\"auber \cite{HowardTauber} 
study various reaction--diffusion systems involving 
the process $A+A\to 0$ and subject to both imaginary and real
noise, the latter being due to coupling to one or more other reaction
processes. They are led to conclude 
that even in the presence of additional real noise 
the imaginary noise cannot be neglected. 

In Sec.\,\ref{secLeqcomplexnoise} we present the master equation and 
equivalent Langevin equation for the process (\ref{2Ato0}).
In Secs.\,\ref{secparity}-\ref{secapproach} 
we address successively various aspects of the
motion in the complex plane.
In Secs.\,\ref{seclambdainfty} and \ref{seclambdazero} 
we consider the 
limiting cases $\lambda\to\infty$ and $\lambda\to 0$, respectively.
In Sec.\,\ref{secnumerical} we investigate the possibility of
alternative ways of simulating reaction--diffusion problems.
We point out several difficulties and questions associated with
simulating the time evolution of the
Poisson variable $\phi$.
Some of the effects of an additional diffusion
term are also briefly discussed.

\section{Langevin equation with imaginary noise}
\label{secLeqcomplexnoise}

For the process (\ref{2Ato0}) the probability $P(n,t)$ of having $n$
particles $A$ at time $t$ obeys the master equation
\begin{eqnarray}
\frac{\dd P(n,t)}{\dd t}&=&\,\half(n+1)(n+2)P(n+2,t)\,-\,
\half n(n-1)P(n,t)
\nonumber\\[2mm]
&&\,+\,\lambda\,\Big[P(n-2,t)\,-\,P(n,t)\Big] \qquad (n=0,1,2,\ldots)
\label{meq}
\end{eqnarray}
with the convention $P(-2,t)=P(-1,t)=0$.
Following Gardiner \cite{Gardiner} one writes for $P(n,t)$ the
Poisson representation
\begin{equation}
P(n,t)\,=\,\int\idd x\iint\idd y\, F(x,y,t)\,p_{x+\ji y}(n)
\label{Poissrepr}
\end{equation}
in which $p_{\phi}(n)=\phi^n\ee^{-\phi}/n!$
is the Poisson distribution of parameter $\phi$.
It is a true probability only for real positive
$\phi$ and a ``quasi-probability''
for arbitrary complex $\phi=x+\ii y$. 
For any $P(n,t)$ there exists a nonunique
$F(x,y,t)$ such that (\ref{Poissrepr}) holds; one may impose 
the additional requirement
that $F$ be real and nonnegative, but even so it is nonunique.
In any case it is normalized such that
$\int\idd x\iint\idd y\, F(x,y,t)=1$. 

The equivalence between the 
Poisson representation method \cite{Gardiner} 
and the second
quantized formalism \cite{Doi,Peliti,LeeCardy} 
was first pointed out by Droz and McKane
\cite{DrozMcKane}, who however avoid discussing Langevin equations.
By either approach one shows that
$P(n,t)$ defined by (\ref{Poissrepr}) satisfies the master equation
(\ref{meq}) if $F(x,y,t)$ is the probablity density of a variable
$\phi=x+\ii y$ obeying the Langevin equation
\begin{equation}
\frac{\dd\phi}{\dd t}=2\lambda-\phi^2\,+\,\sqrt{2\lambda-\phi^2}\,\,\zeta(t)
\qquad [\mbox{ I }]
\label{Leq}
\end{equation}
Here $\zeta(t)$ is Gaussian white noise of autocorrelation
$\langle\zeta(t)\zeta(t')\rangle=\delta(t-t')$ and $[\mbox{ I }]$
indicates the It\^o interpretation \cite{Gardiner,Risken,VanKampen}. 
Equation (\ref{Leq}) 
is the subject of the present study. It derives it interest from
the fact that the noise term is complex
whenever $\phi$ is outside the real interval
$[-\sqrt{2\lambda},\sqrt{2\lambda}]$. 

The field-theoretical approach of Ref.\,\cite{Cardy} 
deals with the case $\lambda=0$
and the noise term in the Langevin equation there appears as
$\xi(t)=\ii\phi(t)\zeta(t)$. Hence
$\xi(t)$ is Gaussian white noise with
\begin{equation}
\langle \xi(t)\xi(t')\rangle=-\phi^2\delta(t-t')
\label{defxi}
\end{equation}
Because of the minus sign on the RHS
of (\ref{defxi}) and perhaps because of the suggestive factor
$-\phi^2$, the noise 
is usually referred to as ``imaginary.''
Of course the noise term as well as the other
terms in the Lange\-vin equation (\ref{Leq}) are generically complex.

The time dependent averages calculated from the  master equation 
(\ref{meq}) are related to those found 
from the Langevin equation (\ref{Leq}) by \cite{Gardiner}
\begin{equation}
\langle\phi^k\rangle=\langle n(n-1)\ldots(n-k+1)\rangle \qquad
(k=0,1,2,\ldots) 
\label{relmoments}
\end{equation}
where the average on the LHS is with respect to $F$ and that on the RHS
with respect to $P$.
Although the average of primary interest, $\langle n\rangle$, is equal
to $\langle\phi\rangle$, it needs to be stressed \cite{Cardy,HowardTauber}
that $\phi$ is not itself a physical
variable. 

{\it Fokker-Planck equation.\,\, }
The Langevin equation (\ref{Leq}), separated into its real and imaginary
parts, is equivalent 
\cite{Gardiner,Risken} to a Fokker-Planck (FP) equation for
the probability density $F(x,y,t)$. We will occasionally refer to
this FP equation, but do not need its explicit general form.
Let us consider, however, the special case of an initial value $\phi(0)$ in
the interval $[-\sqrt{2\lambda},\sqrt{2\lambda}]$. 
Equation (\ref{Leq}) 
shows that then $\phi(t)$ remains confined to this
section of the real axis at all $t>0$. Upon denoting
the time dependent probability density of $\phi=x+\ii y$ 
on this interval
by $F_1(x,t)$ we get from (\ref{Leq})  the equivalent FP equation
\begin{equation}
\frac{\p F_1(x,t)}{\p t}=\Big[\frac{\p}{\p x}-\half\frac{\p^2}{\p x^2}\Big]\,
(2\lambda-x^2)\,F_1(x,t) 
\label{FPeqreal}
\end{equation}
which will be referred to in Secs. \ref{secapproach} and
\ref{seclambdainfty}.

\section{Parity conservation}
\label{secparity}

The master equation (\ref{meq}) leaves the subspaces of even and odd 
$n$ invariant, so that ${\cal P}\equiv\langle(-1)^n\rangle$ 
is a constant of the motion. 
Upon evaluating this average 
by inserting for $P$ its Poisson representation (\ref{Poissrepr}) 
one finds that
\begin{equation}
\langle(-1)^n\rangle=\langle\ee^{-2\phi}\rangle
\label{relcstmotion}
\end{equation}
which shows that $\langle\ee^{-2\phi}\rangle$ 
is the corresponding constant of the motion in $\phi$
language. Within the second
quantized formalism one may arrive at the same relation 
(\ref{relcstmotion})
at the cost of a somewhat greater algebraic effort \cite{vanWijland}.

It is of interest to show explicitly, starting from the Langevin
equation (\ref{Leq}),
that $\langle\ee^{-2\phi}\rangle$ is conserved. 
One may do so by transforming this equation to the new variable 
$\alpha=\ee^{-2\phi}$. 
We recall that the Langevin-It\^o equation 
$\dd\phi/\dd t=G(\phi)+H(\phi)\zeta(t)$ is
equivalent to the Langevin-Stratonovich equation
$\dd\phi/\dd t=G(\phi)-$$\half H(\phi)\dd H/\dd\phi\,+\,H(\phi)\zeta(t)$,
and that in the latter the unknown may be nonlinearly
transformed according to the usual rules of algebra \cite{Gardiner,Risken}.
Hence the Stratonovich 
($[\mbox{\,S\,}]$) 
version of (\ref{Leq}) is
\begin{equation}
\frac{\dd\phi}{\dd t}=2\lambda-\phi(\phi-\half)\,+
\,\sqrt{2\lambda-\phi^2}\,\,\zeta(t)
\qquad [\mbox{ S }]
\label{LeqS}
\end{equation}
which, transformed to an equation for $\alpha$ and reconverted to
It\^o interpretation, yields
\begin{equation}
\frac{\dd}{\dd t}\ee^{-2\phi}=2\ee^{-2\phi}\,
\sqrt{2\lambda-\phi^2}\,\zeta(t)
\qquad [\mbox{ I }]
\label{Leqconserved}
\end{equation}
In the It\^o interpretation the noise $\zeta(t)$ is independent of $\phi(t)$
and so upon averaging (\ref{Leqconserved}) over $\zeta(t)$
the RHS vanishes, after which an average with respect to
$F(x,y,t)$ yields the parity conservation law.

\section{Time evolution: General properties}
\label{sectimeevol}

We first comment on some general properties of the Langevin
equation (\ref{Leq}). These may be analyzed in terms of two sets of
curves in the complex plane, exhibited in Fig.\,1 for the particular
parameter value $\lambda=3$. 

\begin{figure}[tb] 
\begin{center}
    \epsfig{width=8.4cm, height=11cm, angle=-90, file=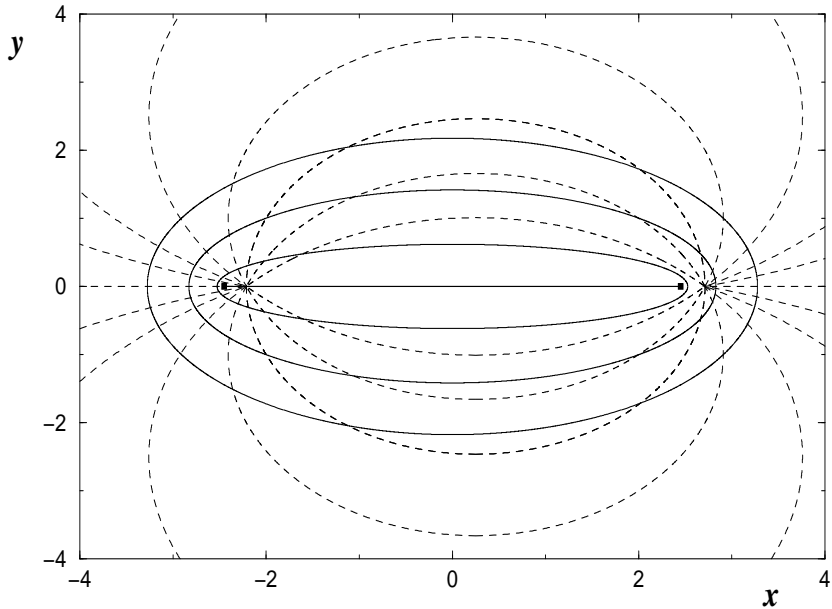}
    \caption{Diffusion curves (solid lines, symmetric about $x=0$) and 
    drift trajectories (dashed lines, symmetric about $x=\frac{1}{4}$) 
    in the complex $\phi=x+\ii y$ plane
    for various values of $C$ and $D$ in Eqs.\,(\ref{setofcurvesC}) 
    and (\ref{setofcurvesD}), respectively, 
    and for $\lambda=3.$ The interval $[-\sqrt{2\lambda},
    \sqrt{2\lambda}]$ is bounded by the two filled squares.
    The drift trajectories all originate
    in a single point $\phi_-$ and terminate in a single point $\phi_+$,
    both on the $x$ axis (see text).}
\end{center}
\end{figure}

{\it Diffusion curves.\,\,}
In any point
of the complex plane the diffusive displacement 
({\it i.e.} the displacement due to
the noise term in (\ref{Leq})) is along a single direction. 
This means that the diffusion tensor in the equivalent FP equation
for $F(x,y,t)$ has one zero
eigenvalue, and therefore this FP equation is not generic.
The zero eigenvalue becomes explicit in an
appropriate set of coordinates, as we will show now.

Upon dividing (\ref{LeqS}) by $\sqrt{2\lambda-\phi^2}$
one finds 
\begin{equation}
\frac{\dd}{\dd t}\,\arcsin\frac{\phi}{\sqrt{2\lambda}}
=\frac{2\lambda-\phi(\phi-\half)}{\sqrt{2\lambda-\phi^2}}\,+\,\zeta(t)
\label{Leqtransformed}
\end{equation}
Then setting \,$\arcsin(\phi/\sqrt{2\lambda})\equiv {\cal Q}(x,y)-\ii 
{\cal R}(x,y)$\,
and taking real and imaginary parts of (\ref{Leqtransformed}) one
gets an expression for $\dd {\cal Q}/\dd t$ which contains the noise 
$\zeta(t)$ and one for $\dd {\cal R}/\dd t$ which shows that ${\cal R}$
is unaffected by the noise. 
The diffusion therefore operates only along curves
${\cal R}(x,y)=C$, where $C$ is a constant. From
the explicit expression
of ${\cal R}(x,y)$ one finds that this collection of
``diffusion curves'' is given by
\begin{equation}
\Big|\,\frac{1-\sqrt{1-2\lambda/\phi^2}}
{1+\sqrt{1-2\lambda/\phi^2}}\,\Big|
\,=\,\ee^{-2C} \qquad (C\geq 0)
\label{setofcurvesC}
\end{equation}
The curves, shown in Fig.\,1, 
are symmetric under reflection with respect to the $x$
and $y$ axis.
They constitute a nested collection of
closed contours that go around the real interval $[-\sqtl,\sqtl]$; 
there is one such contour through each point of the complex plane.
The degenerate contour $C=0$ coincides with $[-\sqtl,\sqtl]$.

{\it Drift trajectories.\,\,}
A second collection of curves is determined by the solutions of
the drift equations, {\it i.e.} of 
(\ref{Leqtransformed}) [or equivalently (\ref{LeqS})]
with $\zeta$ set equal to zero. We will refer to these curves as the
``drift trajectories.''
They may be parametrized by an integration  constant $D$
and the sign of $y$, and are explicitly given by
\begin{equation}
(x-\mbox{$\frac{1}{4}$})^2+y^2=
\mbox{$\frac{1}{16}$}(1+32\lambda)\,+\,D|y| \qquad
(-\infty<D<\infty)
\label{setofcurvesD}
\end{equation} 
They are symmetric under reflection with respect to 
$x=\frac{1}{4}$ and $y=0$.
For all $D$ they have the same pair of end points
$\phi_{\pm}=\frac{1}{4}(1\pm\sqrt{1+32\lambda)}$
which are fixed points of the drift equations.
Time increases from $\phi_-$ to $\phi_+$ 
There is one drift
trajectory through each point of the complex plane.

{\it Existence of solution.}
The drift trajectory consisting of the half-axis $(-\infty,\phi_-)$ is
exceptional: any initial point $x_0$ on this half-axis
arrives at
$x=-\infty$ in a finite ``blowup'' time $t_0$, 
after which the solution of the
drift equations ceases to exist. 
It is therefore necessary to ask if the solution of the full Langevin
equation (\ref{Leq}) exists for all times.
The same question applies to the equivalent FP equation (with, say, an
initial value $F(x,y,0)$ nonzero only in a finite domain).
We consider proving the existence of $F(x,y,t)$ 
(and of averages such as $\langle\phi\rangle$ and 
$\langle\ee^{-2\phi}\rangle$) 
beyond the blowup time as a difficult open problem.
The physicist's ``proof'' consists of remarking that the trajectory
$\phi(t)=x(t)+\ii y(t)$
generated by the Langevin equation never ends, but it ignores the fact
that the diffusion constant associated with this trajectory may become
arbitrarily large.
In any case, a possible singularity that would appear at the
blowup time would certainly not represent a physical effect, but only be
a property of the particular Poisson representation. We return briefly to
these questions in Sec.\,\ref{seclambdazero} when considering the
special case $\lambda=0$.

\section{Equilibrium}
\label{secequilibrium}

It is easy to verify that the
master equation (\ref{meq}) has two independent equilibrium
solutions, {\it viz.$\!$} Poissonians of parameter $\sqtl$ of which one
is restricted to the even and the other to the odd
nonnegative integers. By superposing these solutions one may write
an arbitrary equilibrium solution $P^{\Ind{eq}}(n)$ as
\begin{equation}
P^{\Ind{eq}}(n)=A\,p_{_{\!\sqrt{2\lambda}}}(n)\,+\,
B\,p_{_{\!-\sqrt{2\lambda}}}(n)
\label{statsol}
\end{equation} 
The coefficients $A$ and $B$ are 
determined by the constant of the motion ${\cal P}$ according to
\begin{equation}
A=\frac{\ee^{2\sqrt{2\lambda}}\,-\,{\cal P}}
{2\sinh2\sqrt{2\lambda}}
\qquad
B=\frac{-\ee^{-2\sqrt{2\lambda}}\,+\,{\cal P}}
{2\sinh 2\sqrt{2\lambda}}
\label{exprAB}
\end{equation}
so that $A+B=1$ as required by normalization.
An obvious way to Poisson represent the
equilibrium solution (\ref{statsol}) is
\begin{equation}
F^{\Ind{eq}}(x,y)\,=\,\delta(y)F_1^{\Ind{eq}}(x)
\label{FdF}
\end{equation}
with
\begin{equation}
F_1^{\Ind{eq}}(x)\,=\,A\,\delta(x-\sqtl)\,+\,B\,\delta(x+\sqtl)
\label{statsolPoisson}
\end{equation}
Two limiting cases merit special consideration.
For $\lambda\to\infty$ the coefficient $B$ vanishes exponentially and
the equilibrium solution (\ref{statsolPoisson}) approaches 
$\delta(x-\sqrt{2\lambda})$.
The limit $\lambda\to 0$ is singular (the two delta peaks in
Eq.\,(\ref{statsolPoisson}) coalesce) and requires separate analysis. 
In the $n$ representation one has in this limit
\begin{equation}
P^{\Ind{eq}}(n)=\half(1+{\cal P})\,\delta_{n0}\,+
\half(1-{\cal P})\,\delta_{n1}
\label{statsol0}
\end{equation}
in agreement with the fact
that for $\lambda=0$ there remains 0 or 1 particle, depending on the
initial state. One possible Poisson representation of 
(\ref{statsol0}) is 
\begin{equation}
F_1^{\Ind{eq}}(x)=\half(1+{\cal P})\,\delta(x)\,-\,\half(1-{\cal P})\,
\frac{\dd}{\dd x}\delta(x) 
\label{ststlambda0}
\end{equation}
as is easily verified by substitution in Eq.\,(\ref{Poissrepr}).
The second term in Eq.\,(\ref{ststlambda0}) is no longer nonnegative,
and hence in this case $F_1^{\Ind{eq}}$ is actually a quasi-probability
distribution. 
We emphasize that the Poisson representations exhibited here
are by no means unique.
Below we will see how other representations of the equilibrium
distributions arise.

\section{Approach of equilibrium}
\label{secapproach}

The approach of equilibrium in the $n$ representation is very
unsurprising. For $t\to\infty$ the distribution $P(n,t)$
will tend to an equilibrium $P^{\Ind{eq}}(n)$ 
uniquely determined by the initial value of ${\cal P}$ and
given by (\ref{statsol})--(\ref{exprAB}).  

We will now consider the approach to equilibrium
in the $\phi$ representation for an
initial state $\phi(0)=\rho$, {\it i.e.$\!$} for an initial Poisson 
distribution $P(n,0)=p_{\rho}(n)$.
The constant of the motion then has the value
$\langle(-1)^n\rangle=\ee^{-2\rho}$. 
We are led to distinguish two cases.

{\it Real solutions.}\,\,
\label{secrealsol}
Let us first suppose that the initial value $\rho$ is in
the interval $[-\sqrt{2\lambda},\sqrt{2\lambda}]$. 
In this case
the time dependent probability distribution $F(x,t)$
of $\phi$ on this interval
is described by the equivalent FP equation (\ref{FPeqreal})
For $\lambda>0$ 
the two boundary points are ``adhesive'' (in the terminology of
Ref.\,\cite{VanKampen}). 
For $t\to\infty$ the solution tends to the equilibrium distribution
$F^{\Ind{eq}}(x)$ given by (\ref{statsolPoisson})
and (\ref{exprAB}) with
$\langle(-1)^n\rangle=\ee^{-2\rho}$. 
Note that when $\rho$
is in the interval under consideration, 
we have $A,\,B\geq 0$.

{\it Complex solutions.}\,\,
Let now  $\rho$ be real but outside
$[-\sqrt{2\lambda},\sqrt{2\lambda}]$ (only $\rho>\sqtl$ is physical). 
The Langevin equation (\ref{Leq}) then generates a stochastic 
trajectory $\phi(t)$ in the complex plane.
The time evolution is equivalently described in this case 
by an FP equation for the {\it bivariate}
probability distribution $F(x,y,t)$ with initial condition
$F(x,y,0)=\delta(y)\delta(x-\rho)$.  
Since in this case either $A$ or $B$ is negative, the --~necessarily
nonnegative~-- solution
$F(x,y,t)$ cannot for $t\to\infty$ tend
to the double delta peak given by (\ref{FdF})--(\ref{statsolPoisson}).
Hence something else must happen. 
In fact, for $t\to\infty$ the distribution $F(x,y,t)$ tends to
a stationary distribution $F^{\Ind{eq}}$ 
that has its support in the complex plane. 
To obtain this distribution, we have let a set (a ``cloud'') of 10\,000
points, initially all concentrated in $\phi=1$, evolve independently
in time, 
until their spatial distribution
had reached (or, at least, closely approached) a stationary state, that
we identify with $F^{\Ind{eq}}$. The resulting final cloud is shown in
Fig.\,2 for the special cases
$\lambda=0$ and $\lambda\to\infty$, where for the latter
we have shifted the origin according to
$x=\sqrt{2\lambda}+u$. The distributions $F^{\Ind{eq}}$
for $0<\lambda<\infty$ interpolate smoothly between these two limiting
cases. 
They are Poisson representations of (\ref{statsol}) distinct from
(\ref{statsolPoisson}).
To our knowledge this is the first time that such equilibrium
distributions in the complex plane have been explicitly determined
numerically.  
Because of the initial condition
the distributions shown here are characterized by the constants of the
motion $\langle\ee^{-2\phi}\rangle=\ee^{-2}$ 
and $\langle\ee^{-2\chi}\rangle=\ee^{-2}$. However, different values of
the constants of the motion yield almost identical ({\it i.e.} visually
indistinguishable) distributions; this is because the value
of the constant of the
motion may be changed arbitrarily by a slight redistribution of the
probability density faraway in the left half plane.


\begin{figure}[ht] 
\makebox[7.4cm][l]    
    {\epsfig{width=6.1cm, height=6.7cm, angle=-90, file=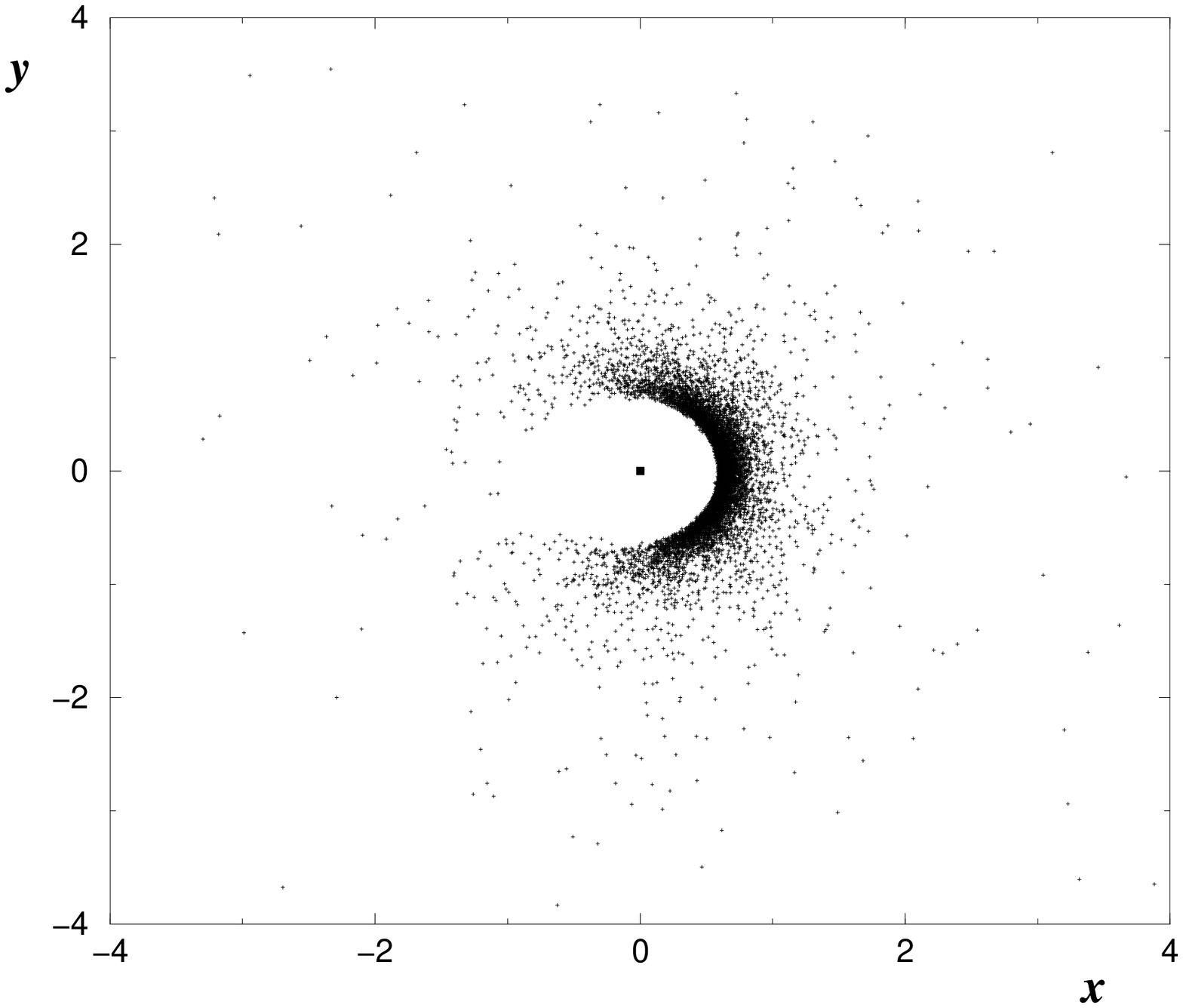}}
\makebox[7.4cm][r]
    {\epsfig{width=6.1cm, height=6.7cm, angle=-90, file=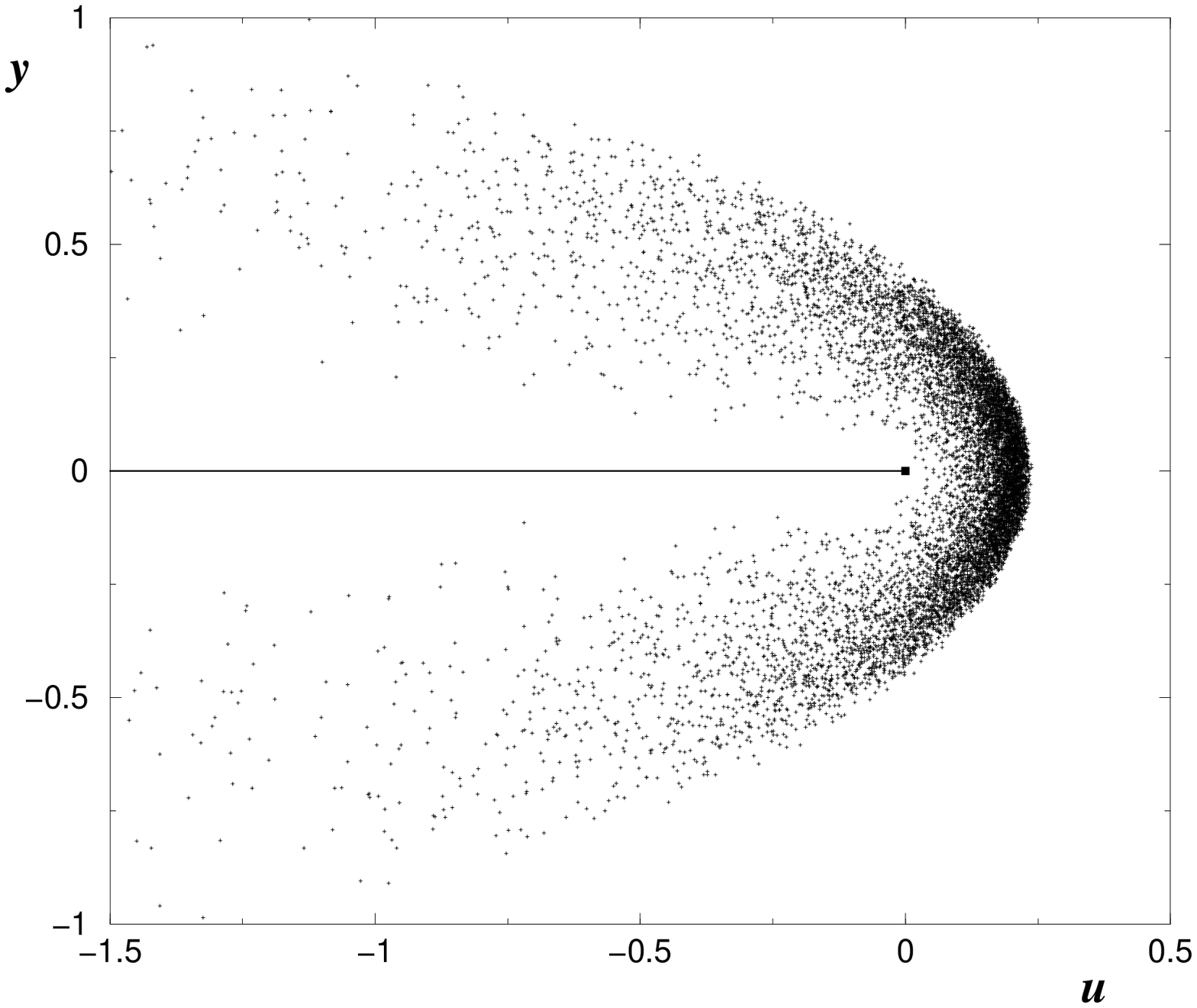}}
\begin{center}
    \caption{{\it Left:\,\,} Cloud of $10\,000$ points representing the 
    stationary probability distribution $F^{\Ind{eq}}(x,y)$
    in the complex plane $\phi=x+\ii y$ for $\lambda=0$.
    {\it Right:\,\,} Cloud of $10\,000$ points representing $F^{\Ind{eq}}(u,y)$
    in the complex plane $\chi=u+\ii y$ for $\lambda=\infty$.} 
\end{center}
\end{figure}
\vspace{-0.8cm}


\vspace{20mm}
\section{Limit $\lambda\to\infty$}
\label{seclambdainfty}

We set $\phi=\sqtl+\chi$ with $\chi=\sqtl+u+\ii y$ 
and scale time according to
$\tau=\sqtl\, t$. The limit $\lambda\to\infty$ of (\ref{Leq}) then
exists and $\chi$ satisfies the Lange\-vin equation
\begin{equation}
\frac{\dd\chi}{\dd\tau}=-2\chi+\sqrt{-2\chi}\,\zeta(\tau) 
\qquad [\mbox{ I }]
\label{Leqchi}
\end{equation}
where $\langle\zeta(\tau)\zeta(\tau')\rangle=\delta(\tau-\tau')$.
This equation conserves the average $\langle\ee^{-2\chi}\rangle$. 
It is interesting to remark that in this limit the drift equation has
become linear so that the problem of blowup in
finite time, discussed in Sec.\,\ref{sectimeevol}, has disappeared.
Below we will 
continue to use the
symbols $F_1$ and $F$ for probability densities on the $u$ axis and
in the $uy$ plane, respectively. 

{\it Real solutions.}
When $\chi$ is real and negative, we merely have
a limiting case of the real problem discussed
in Sec.\,\ref{secapproach}, 
but with the advantage that $F_1(u,t)$ may be
found exactly for arbitrary initial condition $u_0$.
After obtaining and solving an equation for the Fourier transform
of $F_1$ one finds that $F_1(u,t)=
\exp(\frac{2u_0}{\ee^{2t}-1})\delta(u) 
+F_{\Ind{cont}}(u,t)$. Here $F_{\Ind{cont}}(u,t)$ 
is an explicitly known continuous density on the negative
$u$ axis which tends to zero with time for all $u$,
while evolving in such a manner that $\langle\ee^{-2\chi}\rangle$
remains constant. A similar example of such a
solution was worked out in detail in Ref.\,\cite{VanKampen}, ch.\,XII.5.

{\it Complex solutions.}
Any distribution $F$ not initially confined to the negative $u$ axis 
spreads out into the complex $uy$ plane. 
It is then of advantage to transform to the new coordinates 
$R=\mbox{Re}\sqrt{4\chi}$ and $ Q=\mbox{Im}\sqrt{4\chi}$, where $R\geq
0$ and $-\infty< Q<\infty$. 
Up to inessential coefficients these variables
are equal to the $\lambda\to\infty$ limits
of, respectively, ${\cal R}$
and ${\cal Q}$ defined in Sec.\,\ref{sectimeevol}.
In terms of $R$ and $ Q$ the Langevin equation (\ref{Leqchi}) reads
\begin{eqnarray}
\frac{\dd R}{\dd\tau}&=&-R+\frac{R}{R^2+Q^2}
\nonumber\\[2mm]
\frac{\dd Q}{\dd\tau}&=&-Q-\frac{Q}{R^2+Q^2}+\sqrt{2}\,\zeta(\tau) 
\label{LeqRPsi}
\end{eqnarray}
which shows that diffusion takes place only parallel to the $Q$ axis.
The lines $R=E$, where $E$ is a constant, 
correspond to the set of parabolas $u=-y^2/E^2+\frac{1}{4}E^2$ 
in the $uy$ plane.
In spite of this simplification we are not in this case able to 
explicitly find
the stationary density $F^{\Ind{eq}}(u,y)$, 
let alone a time dependent solution $F(u,y,t)$.
Nevertheless, one sees directly from (\ref{LeqRPsi}) 
that $\dd R/\dd \tau<0$ outside the
circle $Q^2+R^2=1$. 
Therefore the probability flow across the line $R=1$ can only be towards
the left and the stationary probability density $F^{\Ind{eq}}(u,y)$ 
must be 
identically zero for $R>1$, {\it i.e.} for $u+y^2>\frac{1}{4}$.

\section{Limit $\lambda\to 0$}
\label{seclambdazero}

For $\lambda\to 0$ the interval $[-\sqtl,\sqtl]$ 
on which there exists a real solution,
contracts to the origin.
When $\lambda=0$ we may set
$\phi=r\ee^{\ji\psi}$ and rewrite the
Langevin equation (\ref{Leq}) in terms of 
the polar coordinates,
\begin{eqnarray}
\frac{\dd r}{\dd t}&=&\,\half r-r^2\cos\psi\nonumber\\[2mm]
\frac{\dd\psi}{\dd t}&=&-r\sin\psi\,+\,\zeta(t)
\label{Langevinrpsi}
\end{eqnarray}
This shows that diffusion takes place only along the angular direction
in the $\phi$ plane. The origin is a fixed point.
Furthermore, since $\dd r/\dd t>0$ inside the
punctuated disk $0<r<\frac{1}{2}$, the probability that flows out of this
disk cannot reenter it, and the stationary probability
density $F^{\Ind{eq}}(x,y)$, shown in Fig.\,3, must be identically zero for
$0<r<\frac{1}{2}$. We remark parenthetically that for $0<\lambda<\infty$
there are no such regions where $F^{\Ind{eq}}$ vanishes.

For $\lambda=0$ the Langevin equation (\ref{Leq})
becomes linear in terms of the variable $\phi^{-1}$. Its solution
with initial value $\phi(0)=\phi_0$ may then be given explicitly and reads
\begin{equation}
\phi(t)=\phi_0\,\Big[\ee^{-\frac{t}{2}-\ji Z(t)}\,
+\,\phi_0 \int_0^t \dd\tau\,\, \ee^{-\frac{\tau}{2}+
\ji Z(t-\tau)-\ji Z(t)}\Big]^{-1} 
\label{solLeqbeta}
\end{equation}
in which $Z(t)$ is the Wiener process
\begin{equation}
Z(t)=\int_0^t \dd\tau\, \zeta(\tau)
\label{defZ}
\end{equation}
The moments $\langle\phi^m(t)\rangle$ may be calculated 
from the $m$th power of expression (\ref{solLeqbeta}), since after
expanding in powers of $\phi_0$ it is possible to average all terms in
the resulting series
explicitly with respect to $Z(t)$.
For $t\to\infty$ the moment $\langle\phi^m(t)\rangle$ 
is found to tend to $\half\delta_{m1}(1-\ee^{-2\phi_0})$,
in agreement with what one may conclude by combining equations
(\ref{relmoments}), (\ref{statsol0}), (\ref{exprAB}), and
(\ref{relcstmotion}). 
Although these moments appear to exist, the derivation is formal in that
it ignores the convergence questions of the series, and hence does not
constitute an answer to the existence problem raised at the end of
Sec.\,\ref{sectimeevol}.

\section{Remarks on numerical simulation}
\label{secnumerical}
It is natural to ask if simulating the Langevin equation for
the field $\phi$ has
any advantages over simulating a particle system. 
Numerical integration of the Langevin equation (\ref{Leq}) 
may be carried out on a ``cloud'' of points in the complex $\phi$ plane,
representative of the function $F(x,y,t)$. 
We illustrate below by means of two examples
two different statistical problems that one encounters, depending on the
choice of Poisson representation.

{\it Problems associated with large $|\phi|$.}\,\,
Fig.\,3 shows, for the special case
$\lambda=0$, the time evolution of the
average $\langle\phi\rangle$ and the constant of the motion
$\langle\ee^{-2\phi}\rangle$ associated with a cloud of 
$10\,000$ points initially all concentrated in $\phi=1$. 
For $t>0$ this cloud spreads out and its average position
starts moving. The initial decay of $\langle\phi\rangle$ 
and the constancy of $\langle\ee^{-2\phi}\rangle$ are in full
agreement with what we know analytically. 
After a relatively short time it will begin to
happen that occasionally one of the points
in the cloud moves to large negative values of
$x$, after which it quickly
returns via a large loop to a faraway region
near the positive $x$ axis. These large excursions, when they
begin to occur, dominate and distort the 
averages, which then become erratic.
In Fig.\,3 this happens for $\langle\phi\rangle$ when $t\gsim 2.5$, and for
the more sensitive average $\langle\ee^{-2\phi}\rangle$
when $t\gsim 1$.

For the initial condition $\phi=1$ the average $\langle\phi\rangle$ 
should for $t\to\infty$ tend exponentially to its equilibrium value 
$\langle\phi\rangle=\frac{1}{2}(1-\ee^{-2})$.
In reality, due to accumulating errors, it 
ends up by fluctuating around 
$\frac{1}{2}$. The
reason for this value is that after having made a large loop, the
trajectories $\phi(t)$ arrive at large $x$ values that are close to even
or odd $n$'s with the same probability. Hence the numerical errors cause
a ``leak'' between the two invariant subspaces (``parity violation'')
which redistributes the total probability equally over both. 
We have not in this study attempted to control these numerical errors.


\begin{figure}[tb] 
\begin{center}
    \epsfig{width=6.1cm, height=8cm, angle=-90, file=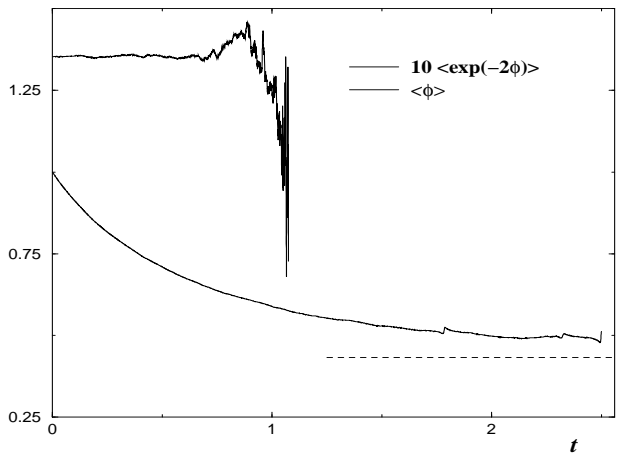}
    \caption{Time evolution of the real part of the average
    $\langle\phi\rangle$ and the 
    conserved parity $\langle\ee^{-2\phi}\rangle=\ee^{-2}$, for an initial
    value $\phi(0)=1$ and for $\lambda=0$. The Monte Carlo average
    is on $10\,000$ realizations of the stochastic process. 
    The dashed line indicates the theoretical equilibrium average
    $\langle\phi\rangle=\frac{1}{2}(1-\ee^{-2})=0.4323\ldots$ 
    for this initial condition.}
\end{center} 
\end{figure}


One ultimate interest is the solution of problems 
with diffusion, {\it i.e.}, for the case with $\lambda=0$, of 
the system of coupled Langevin equations
\begin{equation}
\frac{\dd\phi_j}{\dd t}={\sf D}\,\Delta\phi_j-\phi_j^2
\,+\,\ii\phi_j\,\zeta_j(t)
\qquad [\mbox{ I }]
\label{Leqsystem}
\end{equation}
where ${\sf D}$ is the diffusion constant,
$j$ a lattice site index and $\Delta$ the lattice Laplacian.
The fact that in the Poisson representation
the diffusion process is deterministic,
suggests that Eq.\,(\ref{Leqsystem}) might
be a good starting point for numerical simulation: a part of the 
problem's stochastic character has been eliminated. 
Moreover, upon passing from the single site problem discussed so far
to the lattice problem (\ref{Leqsystem}), 
one might think that the diffusion term, because it tends to equalize the
$\phi_j$, would attenuate the effect of 
excursions deep into the negative half plane. 
Actual simulations show that for large enough ${\sf D}$ indeed it does,
but this advantage is offset by the the fact that the
asymptotic decay of the particle density (well-known to be as
$t^{-1/2}$) then also sets in later.

{\it Problems associated with small $|\phi|$.}\,\,
All Monte Carlo simulations discussed above concerned real positive
probability distributions $F(x,y,t)$.
As remarked in Sec.\,\ref{secLeqcomplexnoise}, a Poisson representation
may also be 
realized with the aid of a quasi-probability, {\it i.e.}
a function $F$ that may take negative (or even complex) values. 
The time evolution of such quasi-probabilities
may also, in principle, be obtained from the Monte Carlo simulation of
the Langevin equation.  We show this on the example of a
state with initially exactly $N$
particles. Such a state
may be Poisson represented by $F(x,y)=\delta(y)F_{1}(x)$ where $F_1(x)$
is the quasi-probability distribution
\begin{equation}
F_{1}(x)=\ee^x\,\Big(\!\!-\frac{\dd}{\dd x}\Big)^{\! N}\,\delta(x)
\label{initialN}
\end{equation}
as is easily verified by substitution in Eq.\,(\ref{Poissrepr}).
(An alternative formula uses a representation on a circle of radius
$\epsilon$ around the origin; we do not discuss this here.)
In order to be able to work numerically with Eq.\,(\ref{initialN}) 
we express the derivative of the Dirac delta as 
\begin{equation}
\frac{\dd}{\dd x}\delta(x)=\frac{1}{\epsilon}\,[\,\delta(x-\half\epsilon)
-\delta(x+\half\epsilon)\,]
\label{ddeltadx}
\end{equation}
where $\epsilon$ should be taken
sufficiently small to have a good approximation.
$N$-fold application of Eq.\,(\ref{ddeltadx}) yields the $N$th derivative as a
sum of $N\!+\!1$ delta functions.
Combining this with Eq.\,(\ref{initialN}) one obtains for the 
special case of $N=4$ 
\begin{equation}
F_{1}(x)=\frac{1}{\epsilon^4}\,[\,\ee^{2\epsilon}\delta(x-2\epsilon)
-4\ee^{\epsilon}\delta(x-\epsilon)+6\delta(x)
-4\ee^{-\epsilon}\delta(x+\epsilon)+\ee^{-2\epsilon}\delta(x+2\epsilon)\,]
\label{initialN4}
\end{equation}
This representation has its support confined within a circle of
arbitrarily small radius around the origin in the complex plane.
We have carried out a simulation of the time evolution of the initial
state (\ref{initialN4})
taking $\epsilon=0.01$ and representing
the initial state by four clouds of 10\,000 points each.
The two located in $x=\pm\epsilon$ each have weight $-4\ee^{\pm\epsilon}$ and 
those in $x=\pm 2\epsilon$ have weight $\ee^{\pm 2\epsilon}$.
Since the FP equation for $F(x,y,t)$
is linear, these weights stay attached to the
clouds during their time evolution.
A trivial cloud with weight $6$ is located in $x=0$; it does not move
with time, so need not be simulated, but enters into the calculation of
averages.

We have taken $\lambda=0$. 
Fig.\,4 shows the decay of the initial state with $N=4$ particles.
The curves do not change when $\epsilon$ is taken smaller.
Again, after a relatively short time strong fluctuations appear and the
result becomes unreliable. This instability is due to the fact that 
the fourth derivative in Eq.\,(\ref{initialN4})
involves the four times repeated subtraction of almost equal numbers.
This explanation is confirmed by the fact that for
$N=2$ the instability appears at a later time, as shown for
comparison in Fig.\,4. 

Another new feature appears in this representation.
Since $\epsilon$ is arbitrarily small and the radial time derivative
$\dd r/\dd t$ is bounded from above, as shown by the first one of
Eqs.\,(\ref{Langevinrpsi}), the clouds of points
will reach the circle $r=\frac{1}{2}$
only after a time $T_\epsilon$ that diverges as $\epsilon\to 0$,
{\it i.e.}, after
the process has come arbitrarily close to equilibrium. 
Hence in the limit $\epsilon\to 0$ the decay to the physical
equilibrium 
takes place inside the disk $|\phi(t)|<\frac{1}{2}$, even though
the equilibrium 
distribution $P^{\Ind{eq}}(n)$ is represented
by a {\it nonstationary}  $F(x,y,t)$ inside this disk.
This simulation is probably closest to the analytical treatment of field
theory, which amounts to working with $\phi$ close to zero.


\begin{figure}[htb] 
\begin{center}
    \epsfig{width=6.1cm, height=8cm, angle=-90, file=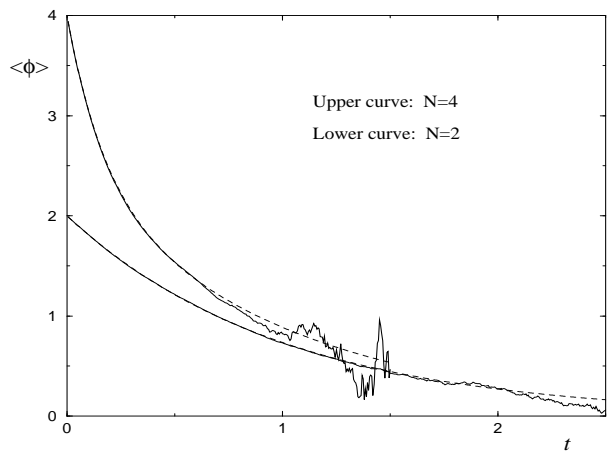}
    \caption{Time evolution of the Monte Carlo averages 
    $\langle\phi\rangle$ for
    initial states with $N=4$ and $N=2$ particles. 
    The dashed curves
    represent the exact solutions.}
\end{center} 
\end{figure}
\vspace{1cm}


\section{Final comments}
\label{secfinal}
We have performed a case study -- to our knowledge 
the first of its kind -- of
various analytical and simulational aspects of 
reaction--diffusion processes in the Poisson representation.
We have identified many curious and interesting phenomena
that happen to the probability flow in this representation. 
This study is an exploration, and necessarily far from exhaustive. We
have not discussed, for example, 
the equilibrium fluctuations that occur when $\lambda>0$.
One of
our motivations was the search for different and possibly more efficient
Monte Carlo simulation methods for such processes. It seems clear that
such a hope is not easily realized. 
A more speculative perspective
is a combination of the present kind of
Monte Carlo simulation
with renormalization.

\section*{Acknowledgments} 
The authors 
have benefitted from discussions with F. van Wijland
and from correspondence with M. Droz.



\end{document}